\documentclass[journal]{IEEEtran}

\usepackage{bm}
\usepackage{color}

\usepackage{soul}

\usepackage[noadjust]{cite}

\usepackage[rightcaption]{sidecap}

\ifCLASSINFOpdf

  \usepackage{gensymb}

   \usepackage[pdftex]{graphicx}

   \usepackage{multirow}

   \usepackage{tabularew}

   \usepackage[flushleft]{threeparttable}

   \graphicspath{{./pdf/}{./jpeg/}{./eps/}}

  \DeclareGraphicsExtensions{ ,.jpeg,.png}

\else

   \usepackage[dvips]{graphicx}

   \graphicspath{{./eps/}}

   \DeclareGraphicsExtensions{.eps}

\fi

\usepackage{amsmath}

\usepackage{array}

\ifCLASSOPTIONcompsoc

\usepackage[caption=false,font=normalsize,labelfont=sf,textfont=sf]{subfig}

\else

  \usepackage[caption=false,font=footnotesize]{subfig}

\fi

\usepackage{stfloats}

\usepackage{url}

  \usepackage{color}

\usepackage{url}

\begin{document}



\title{Intrinsic Spike Timing Dependent Plasticity in Stochastic Magnetic Tunnel Junctions Mediated by Heat Dynamics}

\author{Humberto Inzunza Velarde*, Jheel Nagaria*, Zihan Yin, Ajey Jacob, Akhilesh~Jaiswal

\thanks{The authors are with the Information Sciences Institute, University of Southern California, Los Angeles, CA-90007, USA. email: inzunzav@usc.edu, nagaria@usc.edu, zihanyin@usc.edu, ajey@isi.edu, akjaiswal@isi.edu}

\thanks{(* These authors contributed equally)}

}


\maketitle

\pagenumbering{gobble}

\begin{abstract}

The quest for highly efficient cognitive computing has led to extensive research interest for the field of \textit{neuromorphic computing}. Neuromorphic computing aims to mimic the behavior of biological neurons and synapses using solid-state devices and circuits. Among various approaches, emerging non-volatile memory technologies are of special interest for mimicking neuro-synaptic behavior. These devices allow the mapping of the rich dynamics of biological neurons and synapses onto their intrinsic device physics. In this letter, we focus on Spike Timing Dependent Plasticity (STDP) behavior of biological synapses and propose a method to implement the STDP behavior in Magnetic Tunnel Junction (MTJ) devices. Specifically, we exploit the time-dependent heat dynamics and the response of an MTJ to the instantaneous temperature to imitate the STDP behavior. Our simulations, based on a macro-spin model for magnetization dynamics, show that, STDP can be imitated in stochastic magnetic tunnel junctions by applying simple voltage waveforms as the spiking response of pre- and post-neurons across an MTJ device.

\end{abstract}


\begin{IEEEkeywords}

Spike Timing Dependent Plasticity, STDP, Magnetic Tunnel Junctions, MTJ, Neuromorphic Computing, Heat Dynamics.

\end{IEEEkeywords}

\IEEEpeerreviewmaketitle

\section{Introduction}

The remarkable energy-efficiency of the biological brain coupled with its cognitive ability has recently attracted considerable research investigations \cite{doi:10.1152/jn.00041.2021,cauwenberghs2013reverse}. In the quest towards highly energy-efficient systems, the field of \textit{neuromorphic computing}, aims at imitating neuro-synaptic dynamics for creating bio-plausible computing hardware \cite{roy2019towards}. Towards that end, CMOS \cite{liu2002cmos,seo201145nm,wu2015cmos} as well as various emerging technologies are being widely explored for implementing neural and synaptic functionalities \cite{jaiswal2017proposal,sengupta2016magnetic,bichler2012visual, jo2010nanoscale}. Of special interest are emerging non-volatile technologies due to their compact size and non-volatility leading to dense integration and power efficiency \cite{nvsim,cite-key,9474064,9365820}. Among emerging non-volatile technologies, Magnetic Tunnel Junction (MTJ) based memory elements posses unique desirable features including, almost unlimited endurance, high read-write speed and power efficient switching \cite{huai2008spin,khvalkovskiy2013basic,stt_endur}. 

MTJs have been used both as a leaky-integrate-fire neural \cite{sengupta2016magnetic} and Spike Timing Dependent Plasticity (STDP) based synaptic devices \cite{srinivasan2016magnetic}. Specifically, the STDP behavior in MTJs has relied on tuning the switching behavior of the MTJs through timing dependent complex waveforms generated from peripheral circuits \cite{doi:10.1098/rsta.2019.0157}. In this paper, we show for the first time, that the intrinsic physics of the MTJ switching behavior, dependent on the current induced Joule heating effect, can be used to map STDP functionality. Furthermore, our proposed `intrinsic STDP’ behavior in MTJs relies on the time-dependent heat accumulation and dissipation dynamics. Thus, it does not require complex timing circuits to induce STDP like behavior externally.

\section{Intrinsic STDP in Stochastic MTJs Using Heat Dynamics}

 \begin{figure}[t]
    \centering
    \includegraphics[width=0.5\textwidth, height = 1.4in]{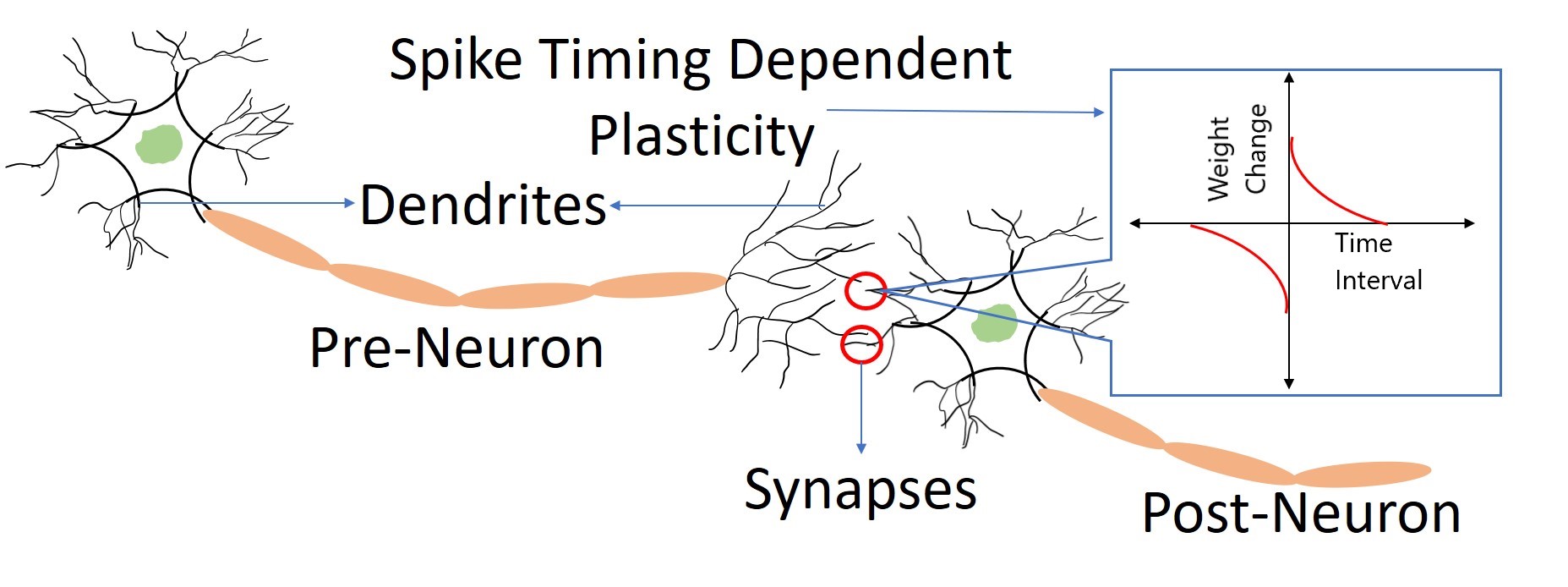}
    \caption{Biological neurons connected through intermediate synapses. Synapses are responsible for the learning process. Spike Timing Dependent Plasticity (STDP) is a biological process wherein the synaptic weights are adjusted based on the precise timing of firing events between the pre- and post- neurons.}
    \label{fig1}
\end{figure}

Spike Timing Dependent Plasticity (STDP) is a phenomenon observed in mammalian brains, wherein the sign and magnitude of the changes in synaptic weights is dependent on the precise timing of the spikes generated from a pre-neuron and a post-neuron \cite{doi:10.1146/annurev.neuro.31.060407.125639}. Specifically, referring to Fig. \ref{fig1}, the pre-neuron and the post-neuron are connected to each other through synaptic junctions between their respective dendrites. When the pre-neuron fires (or emits a spike or action potential) before the post-neuron, the synaptic weight of the concerned synapse is incremented. Furthermore, the magnitude of the increment is proportional to the proximity in time of the spiking events of the pre- and the post-neurons. Conversely, when the post-neuron fires before the pre-neuron, the weight of the intermediate synapse is decremented according to the proximity of the firing event of the post- and the pre-neuron. STDP is considered to be the underlying learning mechanism in the biological brain that leads to long-term changes in synaptic weights based on spike timings \cite{DAMOUR2015514}.

\begin{figure*}[t]
    \centering
    \includegraphics[width= .8\textwidth]{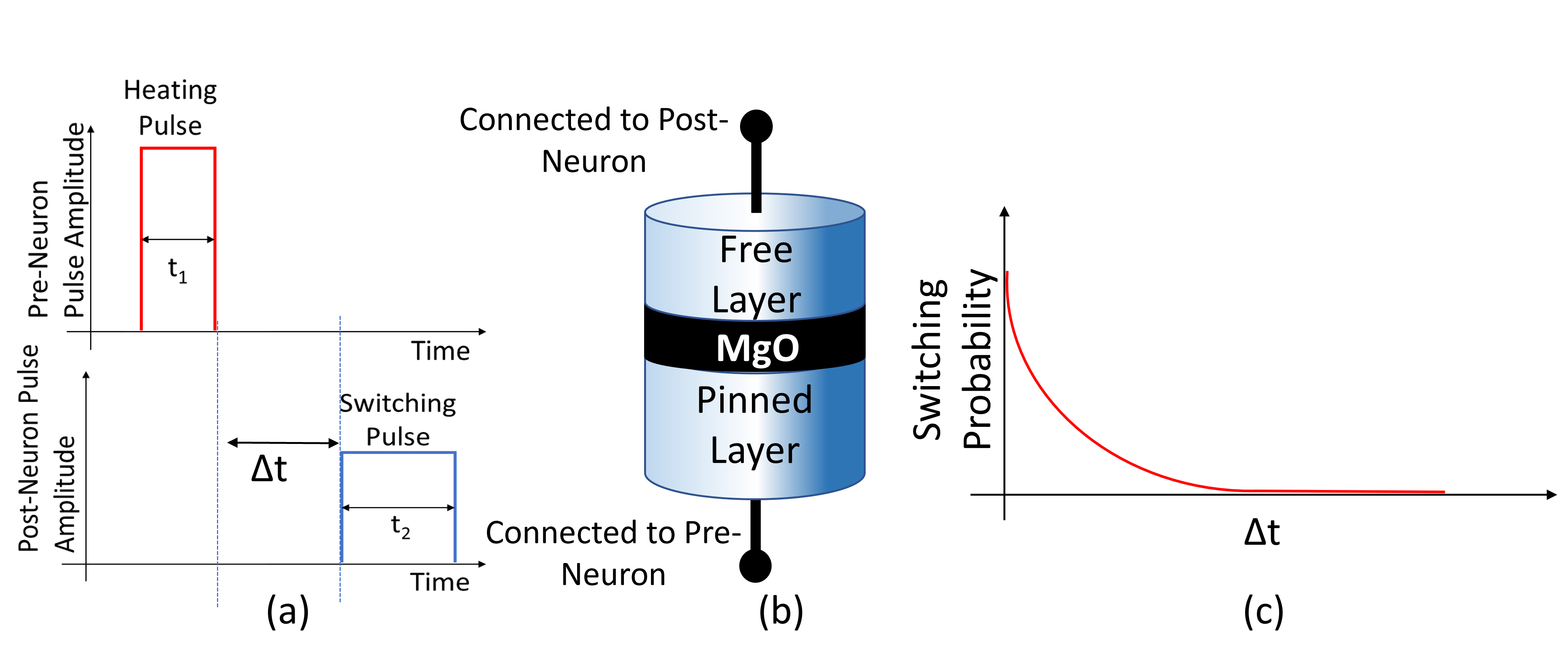}
    \caption{(a) Figure showing the parameters associated with the heating and the switching pulses. (b) A magnetic tunnel junction device comprising of two nano-magnets separated by a non-magnetic spacer (MgO). (c) The expected switching probability behavior of the MTJ due to the pulses shown in part (a).}
    \label{fig2}
\end{figure*}

From a device perspective, the long-term changes in weights can be mapped to the stable resistance states of an MTJ. An MTJ consists of two ferromagnetic layers that sandwich a non-magnetic insulator layer. One of the two ferromagnetic layers, called the \textit{free layer}, can be switched such that its magnetization vector is either parallel or anti-parallel to the \textit{pinned layer}. In the \textit{parallel or (P)} state, the resistance of the MTJ is low as compared to the \textit{anti-parallel or (AP)} state. Thus, the two states - P and AP,  of an MTJ can be used to represent a two-level weight of a synaptic connection. STDP with such MTJs can be implemented by making the switching probability of the MTJs a function of timings of the spikes arriving on its two terminals. Specifically, suppose a spike arrives first on the terminal of the MTJ connected to the pre-neuron, followed by a spike on the second terminal connected to the post-neuron. In such a scenario, the MTJ switches probabilistically from high resistance to low-resistance state, such that the switching probability follows the typical STDP curve for weight increment (MTJ conductance increment). Conversely, when the terminal connected to the post-neuron receives a spike first, followed by the pre-neuron, the MTJ switches probabilistically from low-resistance to high-resistance state, following the STDP curve, leading to a decrease of synaptic weight (MTJ conductance).

To see how such STDP behavior can be obtained intrinsically for the switching probability of an MTJ, let us consider Fig. \ref{fig2}. In Fig. \ref{fig2}, the pinned layer terminal of the MTJ is connected to the pre-neuron, while the free layer terminal is connected to the post-neuron. Initially, when the pre- and the post-neuron are idle (i.e., not spiking), the voltage at the respective MTJ terminals is at 0V. When the pre-neuron fires, it applies a voltage pulse across the MTJ of pulse duration $t_1$. This pulse is a high-current pulse, and as the current flows through the MTJ, it generates heat due to Joule effect. Thus, the \textit{heating pulse}, shown in Fig. \ref{fig2}(a), heats the MTJ above its nominal temperature. Note, the pulse width of this pulse can be chosen such that the MTJ does not switch due to the heating pulse. For example, the heating pulse duration can be smaller than the incubation delay of the MTJ, thus ensuring the MTJ does not have sufficient time to undergo switching \cite{devolder2008single}. As a result, the instantaneous temperature of the MTJ increases due to the heating pulse. Subsequently, the Joule heat generated due to the heating pulse slowly starts to dissipate over time. This, in turn, gradually decreases the temperature of the MTJ over a period of time. After the heating pulse has been applied, let us assume the post-neuron spikes after an interval of $\Delta t$. When the post neuron spikes, it applies a switching pulse of duration $t_2$ on the lower terminal of the MTJ. The switching pulse would switch the MTJ based on the instantaneous temperature of the MTJ. It is well-known that
MTJ switching probability is a strong function of thermal agitation due to their nanoscopic sizes. Therefore, the higher the temperature, higher is the switching probability \cite{jaiswal2016comprehensive}.

\begin{figure*}[t]
    \centering
    \includegraphics[width=1\textwidth, ]{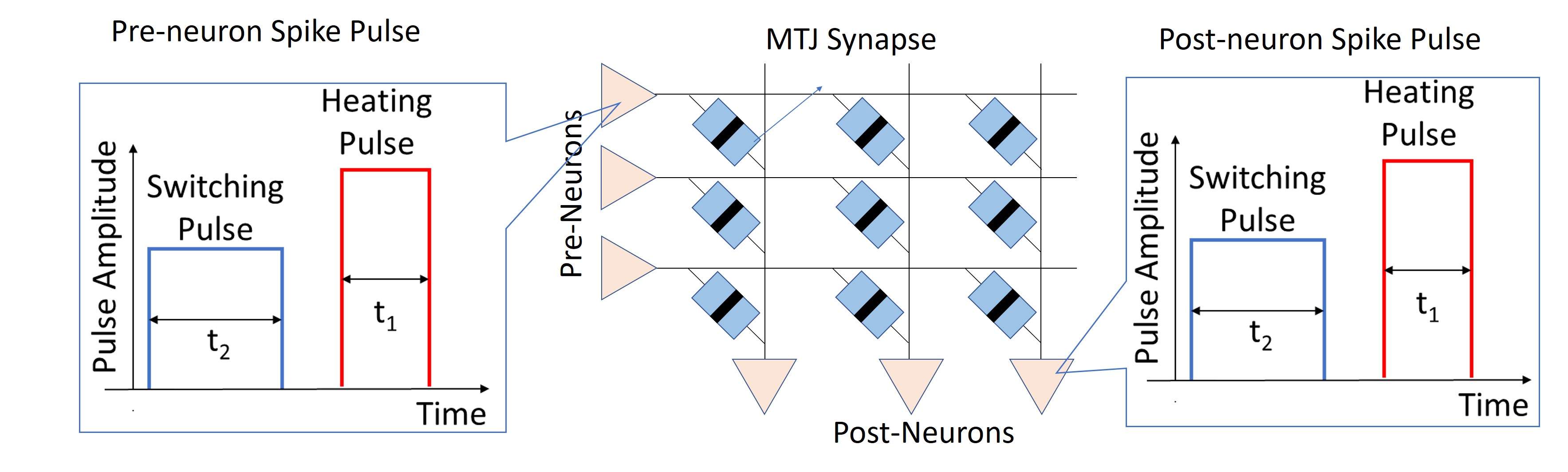}
    \caption{MTJs arranged in a crossbar fashion between the pre- and post-neurons. Note, the heating and the switching pulse combination ensures that a particular MTJ experiences heating pulse from one set of neurons (pre- or post) and a switching pulse from another set or neuron (post or pre). }
    \label{fig3}
\end{figure*}

In other words, the pre-neuron increases the instantaneous temperature of the MTJ, which then decays over time. When the post-neuron spikes, it applies a switching pulse (usually of a lower magnitude and larger pulse duration) in an attempt to switch the MTJ. The resulting probability of switching depends on the instantaneous MTJ temperature. The closer in time the post-neuron spikes with respect to the pre-neuron, higher is the MTJ temperature and hence, higher is the switching probability. As shown, in Fig. \ref{fig2}(c), the switching probability of the MTJ, in presence of the heating and the switching pulse decays as the time interval $\Delta t$ increases between the heating and the switching pulse. The switching behavior shown in Fig. \ref{fig2}(c) resembles the STDP behavior of Fig. \ref{fig1}, for the first quadrant. A similar approach can be used to mimic the switching behavior of the STDP curve for the third quadrant by applying a heating and switching pulse to switch the MTJ in the opposite direction. 

Now, let us consider constructing a crossbar array using MTJ devices exhibiting the intrinsic STDP behavior as described above. Fig. \ref{fig3} shows MTJs arranged in a crossbar fashion. For ease of representation, the selector devices in series with each MTJ are not shown in the figure. The horizontal lines are connected to pre-neurons, while the vertical lines are connected to post-neurons. Thus, the pinned layer (free layer) terminal of each MTJ is connected to a pre- (post- ) neuron. The voltage pulses that are generated by the pre- and the post-neurons, when they emit a spike is represented in Fig. \ref{fig3}(a) and (c), respectively. The first pulse, shown in blue, is the switching pulse for the respective neurons, followed by a heating pulse. This arrangement ensures, when looking at the firing event of the pre-neuron, there exists a time $\Delta t$ between the heating pulse of the pre-neuron and the switching pulse of the post-neuron. Similarly, if the post-neuron fires first, there would exist a time $\Delta t$ between the heating pulse of the post-neuron followed by the switching pulse of the pre-neuron. Further, the pulse associated with the pre-neuron  and post-neuron ensures the weight of the MTJ is probabilistically incremented (decremented) when the pre-neuron (post-neuron) fires first, followed by the post-neuron (pre-neuron). Thus, the intrinsic heat dynamics of the MTJ can be used to construct a crossbar array of MTJs exhibiting probabilistic STDP behavior.

\section{Modeling Framework}

The evolution of the magnetization vector over time was modeled using the Landau-Lifshitz-Gilbert-Slonczewski (LLG-S) equation (Equation 1) under the macro-spin approximation \cite{jaiswal2016comprehensive}. 
\begin{eqnarray}
\frac{d\widehat{m}}{dt}=-\gamma (\widehat{m} \times H_{eff}) + \alpha (\widehat{m} \times \frac{d\widehat{m}}{dt})\nonumber \\
+ \frac{1}{q{N_{s}}}(\widehat{m} \times \widehat{m} \times I_{s}\widehat{M} )
\end{eqnarray}
where, $I_{s}$ is the spin current flowing through the MTJ and represent the effect of spin transfer torque, $\widehat{M}$ is magnetization direction of PL. $\widehat{m}$ represents a unit vector in the direction of magnetization of the free-layer, $\gamma$ represents the gyromagnetic ratio for an electron, $\alpha$ is the Gilbert damping constant, $H_{eff}$ is the effective magnetic field including the demagnetization anisotropy field \cite{Mart_nez_Huerta_2013} and interface anisotropy field \cite{doi:10.1063/1.3670002}, and the effective field due to thermal noise. $N_{s}=M_{s}V/\mu_{B}$ is the number of spins in the free layer volume $V$ ($M_{s}$ is saturation magnetization and $\mu_{B}$ is Bohr magneton). Thermal noise and its effect on the transient response of the MTJ, the Joule heating effect along with the thermal distribution of the initial angle of switching was included in the model using the following equations \cite{papusoi2008probing, jaiswal2016comprehensive}, where $dt$ is simulation timestep, $M_s$ is the saturation magnetization, $V$ is volume, $T$ is temperature, $P_{HP}$ is power supplied to the MTJ, $t$ is time, the subscript $RT$ refers to room temperature, ${\alpha}_J$ is Joule heating constant, $\tau_{TR}$, is characteristic time constant, $T_{AF}$ is temperature of the free-layer, $T_0$ is the initial temperature of the free-layer, $\zeta$ is a random vector with standard deviation of 1. Equation 2 is the heating equation, Equation 3 represents the dynamics of MTJ cooling in absence of heating pulse and Equation 4 is the effective thermal field experienced by the free-layer.

\begin{eqnarray}
T_{AF} = T_{RT} + {\alpha}_J P_{HP} \left( 1 - e^{\frac{-t}{\tau_{TR}}} \right)
\end{eqnarray}

\begin{eqnarray}
T_{AF} = T_0 + (T - T_0 )\left( 1 - e^{\frac{-t}{\tau_{TR}}} \right)
\end{eqnarray}


\begin{eqnarray}
H_{therm}&=& \zeta\sqrt[]{\frac{\alpha K T_{AF} dt}{|\gamma| M_s(T) V }}\label{eqn_H_therm}
\end{eqnarray}

\begin{table}[!t]
\renewcommand{\arraystretch}{1.3}
\caption{PARAMETERS USED IN LLGS SIMULATIONS \cite{jaiswal2016comprehensive, papusoi2008probing}}
\label{table_I}
\centering
\begin{tabular}{ccc}
\hline
\hline
\bfseries  & \bfseries Values   \\
\hline
\bfseries Interface Anisotropy ($K_{i}$) $erg/cm^{2}$ &	\bfseries 1.3	\\
\bfseries Saturation Magnetization ($M_{s}$)  $emu/cm^{3}$ &	\bfseries 1257.3	\\
\bfseries Damping factor ($\alpha$) & \bfseries 0.015 \\
\bfseries Energy Barrier ($E_{B}$) $KT$ &	\bfseries 73 \\
\bfseries Aspect Ratio ($AR$) &	\bfseries 1	\\
\bfseries Free Layer Minimum Diameter $nm$	& \bfseries 40	\\
\bfseries Free Layer Thickness ($t_{fl}$) $nm$ &	\bfseries 1.42 \\

\bfseries Room Temperature ($T_{RT}$) $K$ & \bfseries 300 \\
\bfseries Joule Heating Constant (${\alpha}_J$) $K/W$ &  \bfseries 83600   \\
\hline
\hline
\end{tabular}
\end{table}

\section{Results and Discussions}

\begin{figure}[t]
    \centering
    \includegraphics[width= .45\textwidth]{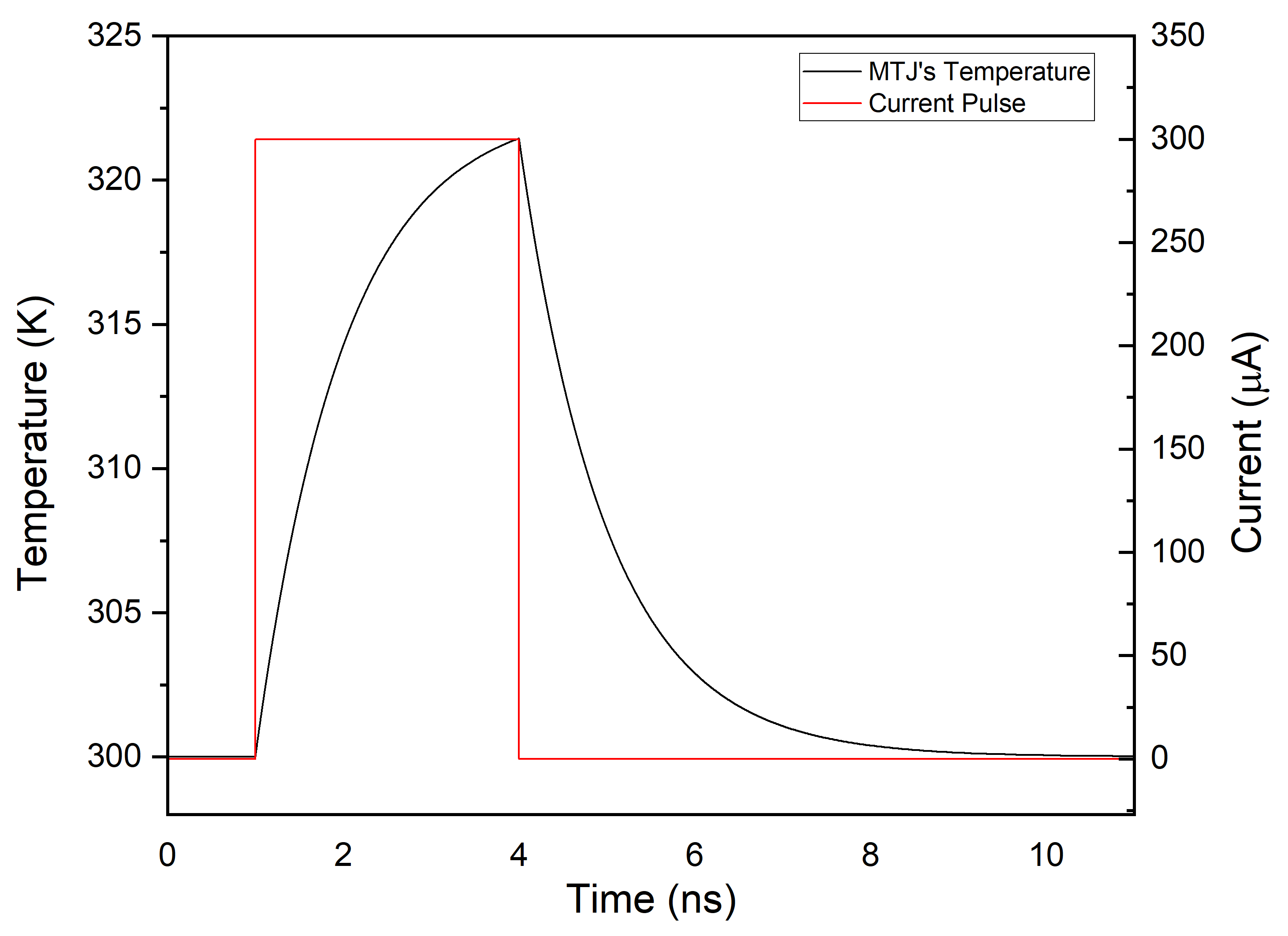}
    \caption{The temperature dynamics of the MTJ in presence of a heating pulse.}
    \label{fig4}
\end{figure}

Fig. \ref{fig4} shows the increase in temperature of the MTJ due application of a voltage pulse across it. The figure depicts the characteristic heat dynamics of an MTJ as a function of time. This heat dynamics results in the intrinsic STDP behavior within MTJs, when connected in a crossbar fashion, as shown in Fig. \ref{fig3}. The accumulation of heat due to passage of an electric current and the subsequent decay of heat over time is reminiscent of the accumulative and dissipative biochemical processes in the biological brain \cite{dayan2001theoretical}.

\begin{figure}[t]
    \centering
    \includegraphics[width= .4\textwidth]{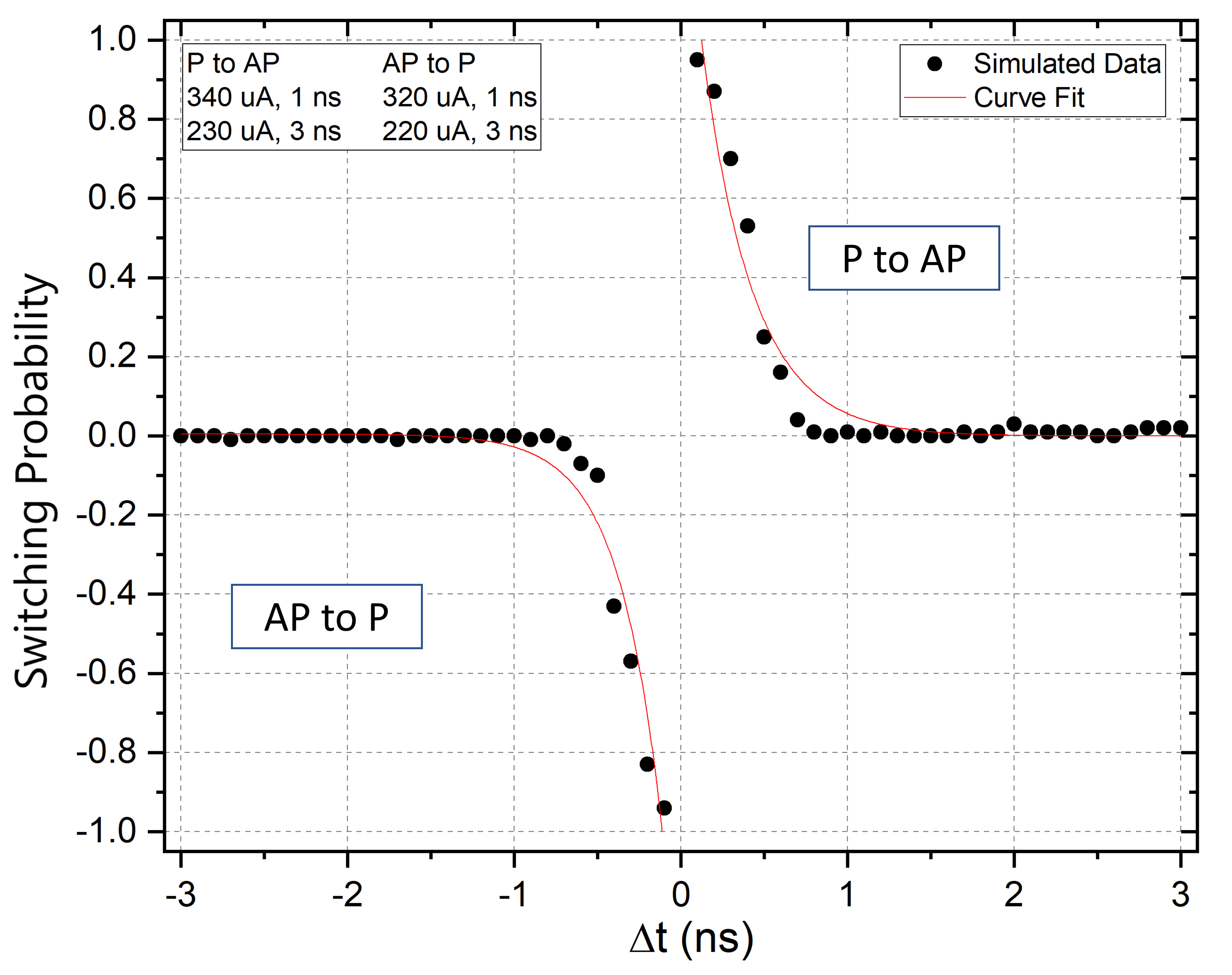}
    \caption{STDP behavior in stochastic MTJs, using a combination of heating and switching pulses. Note, a positive (negative) switching probability indicates switching from AP to P (P to AP) state. In the inset, the upper (lower) number represents magnitude and pulse duration of heating (switching) pulse.}
    \label{fig5}
\end{figure}

Let us now consider Fig. \ref{fig5} that exhibits the desired STDP behavior in accordance with the presented proposal. The x-axis represents the time interval $\Delta t$ between the heating and the switching pulse. A lower $\Delta t$ results in a higher instantaneous temperature of the MTJ and hence, higher switching probability. Further, a positive $\Delta t$ implies the pre-neuron fired before the post-neuron, whereas a negative $\Delta t$ indicates the post-neuron spiked before the pre-neuron. The y-axis represents the magnitude of the switching probability of the MTJ. By the adopted convention, a positive value on the y-axis represents the MTJ is switching from parallel to anti-parallel state. In contrast, a negative value suggests that the switching has occurred in the opposite direction. Thus, a long-term increase or decrease in the synaptic weights occurs automatically based on the timing of the skiping events of pre- and post-neurons. The pulse parameters used for generating the plots are mentioned as an inset in the figure. Thus, Fig. \ref{fig5} represents the intrinsic STDP behavior of the MTJs under heating and switching pulses.

\begin{figure}[t]
    \centering
    \includegraphics[width= .4\textwidth]{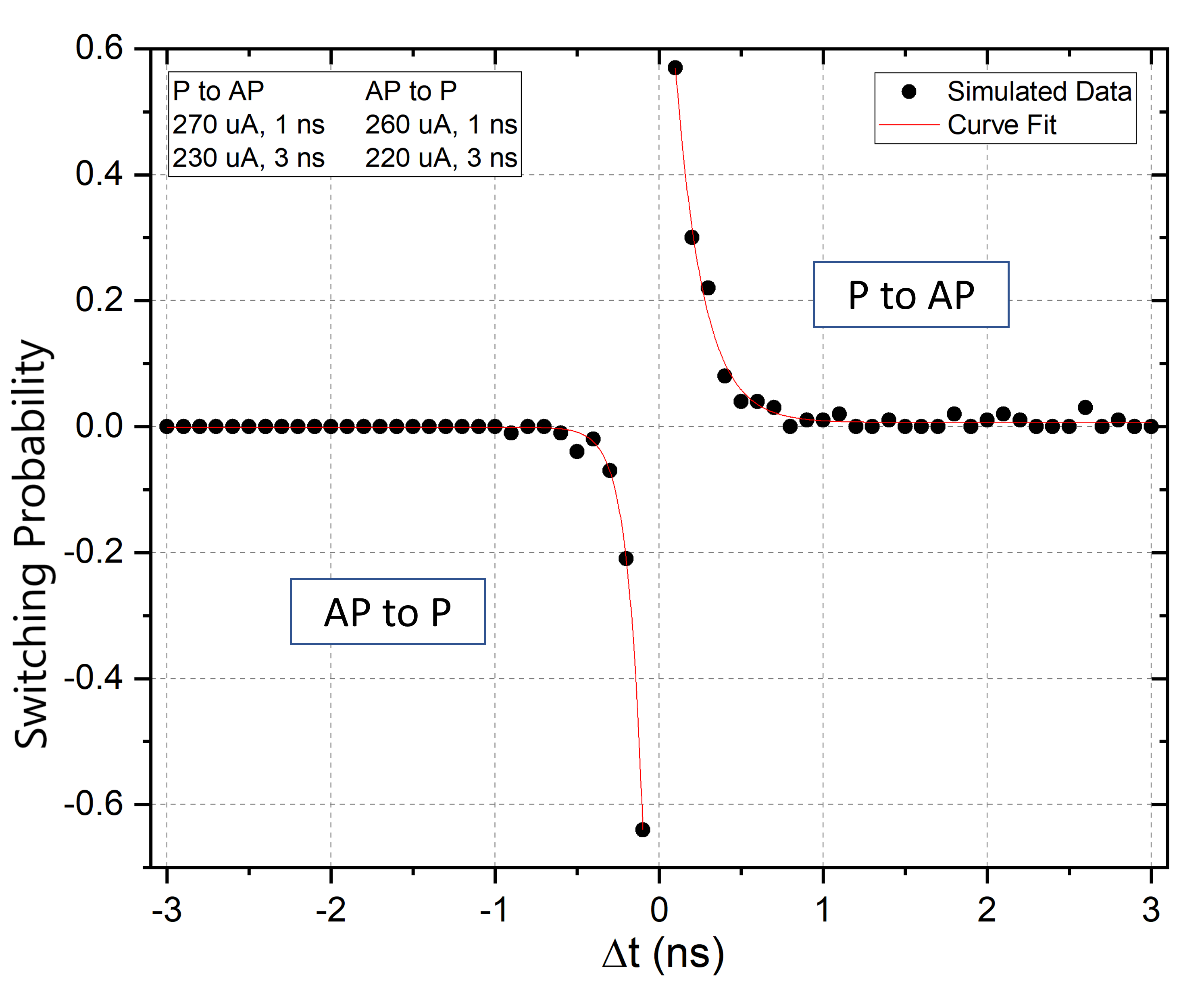}
    \caption{STDP behavior, similar to Fig. \ref{fig4}, wherein the maximum achievable switching probability is adjusted by adjusting the input pulses.  In the inset, the upper (lower) number represents magnitude and pulse duration of heating (switching) pulse}
    \label{fig6}
    \vspace{-.2in}
\end{figure}

With respect to the presented STDP behavior, we would like to highlight few key points. The shape of the STDP curve can be controlled by various electrical, device and material parameters. For example, a lower magnitude of heating or switching pulse can reduce the maximum achievable switching probability for the MTJ STDP behavior. Similarly, increasing the energy barrier of the magnets through device dimensions or material parameters would also affect the magnitude of the switching probability curve. As an example, Fig. \ref{fig6} presents the STDP behavior with a different set of pulse parameters exhibiting the fact that the STDP curve can be engineered by controlling various MTJ parameters. 


\section{Conclusions}

In the quest for energy-efficiency, neuromorphic computing aims at mimicking the neuro-synaptic computations in the human brain. However, the underlying neuro-synaptic biological processes exhibit rich dynamical behavior, which is difficult to mimic using conventional CMOS technology. In this letter, we present a proposal to map the Spike Timing Dependent Plasticity (STDP) behavior of biological synapses through the intrinsic switching characteristics of magnetic tunnel junctions (MTJ). This characteristic switching behavior emerges due to a combination of accumulative and dissipative heat dynamics in conjunction to the effect of thermal noise on the switching behavior of MTJs. Specifically, our proposal consists of application of a heating and a switching pulse to the MTJ. The MTJ being heated, for example by a pre-neuron, switches with a higher switching probability when a switching pulse is applied by the firing event of a post-neuron. The closer in time the switching events of the two neurons, higher is the instantaneous MTJ temperature and hence higher is the switching probability, thus mimicking the STDP behavior. Furthermore, our proposal opens up pathways to further replicate other neuro-synaptic behavior that in general depends on accumulative and dissipative bio-chemical processes. 
 
\section*{Acknowledgement}
This research was supported in part by the Center for Undergraduate Research at Viterbi Engineering (CURVE), University of Southern California, Los Angeles, USA.

\bibliographystyle{IEEEtran}
\bibliography{IEEEabrv,ref}

\begin{thebibliography}{10}
\providecommand{\url}[1]{#1}
\csname url@samestyle\endcsname
\providecommand{\newblock}{\relax}
\providecommand{\bibinfo}[2]{#2}
\providecommand{\BIBentrySTDinterwordspacing}{\spaceskip=0pt\relax}
\providecommand{\BIBentryALTinterwordstretchfactor}{4}
\providecommand{\BIBentryALTinterwordspacing}{\spaceskip=\fontdimen2\font plus
\BIBentryALTinterwordstretchfactor\fontdimen3\font minus
  \fontdimen4\font\relax}
\providecommand{\BIBforeignlanguage}[2]{{%
\expandafter\ifx\csname l@#1\endcsname\relax
\typeout{** WARNING: IEEEtran.bst: No hyphenation pattern has been}%
\typeout{** loaded for the language `#1'. Using the pattern for}%
\typeout{** the default language instead.}%
\else
\language=\csname l@#1\endcsname
\fi
#2}}
\providecommand{\BIBdecl}{\relax}
\BIBdecl

\bibitem{doi:10.1152/jn.00041.2021}
\BIBentryALTinterwordspacing
Z.~Z. Haque, R.~Samandra, and F.~A. Mansouri, ``Neural substrate and underlying
  mechanisms of working memory: insights from brain stimulation studies,''
  \emph{Journal of Neurophysiology}, vol. 125, no.~6, pp. 2038--2053, 2021,
  pMID: 33881914. [Online]. Available:
  \url{https://doi.org/10.1152/jn.00041.2021}
\BIBentrySTDinterwordspacing

\bibitem{cauwenberghs2013reverse}
G.~Cauwenberghs, ``Reverse engineering the cognitive brain,'' \emph{Proceedings
  of the national academy of sciences}, vol. 110, no.~39, pp. 15\,512--15\,513,
  2013.

\bibitem{roy2019towards}
K.~Roy, A.~Jaiswal, and P.~Panda, ``Towards spike-based machine intelligence
  with neuromorphic computing,'' \emph{Nature}, vol. 575, no. 7784, pp.
  607--617, 2019.

\bibitem{liu2002cmos}
B.~Liu and J.~F. Frenzel, ``A cmos neuron for vlsi circuit implementation of
  pulsed neural networks,'' in \emph{IEEE 2002 28th Annual Conference of the
  Industrial Electronics Society. IECON 02}, vol.~4.\hskip 1em plus 0.5em minus
  0.4em\relax IEEE, 2002, pp. 3182--3185.

\bibitem{seo201145nm}
J.-s. Seo, B.~Brezzo, Y.~Liu, B.~D. Parker, S.~K. Esser, R.~K. Montoye,
  B.~Rajendran, J.~A. Tierno, L.~Chang, D.~S. Modha \emph{et~al.}, ``A 45nm
  cmos neuromorphic chip with a scalable architecture for learning in networks
  of spiking neurons,'' in \emph{2011 IEEE Custom Integrated Circuits
  Conference (CICC)}.\hskip 1em plus 0.5em minus 0.4em\relax IEEE, 2011, pp.
  1--4.

\bibitem{wu2015cmos}
X.~Wu, V.~Saxena, and K.~Zhu, ``A cmos spiking neuron for dense
  memristor-synapse connectivity for brain-inspired computing,'' in \emph{2015
  International Joint Conference on Neural Networks (IJCNN)}.\hskip 1em plus
  0.5em minus 0.4em\relax IEEE, 2015, pp. 1--6.

\bibitem{jaiswal2017proposal}
A.~Jaiswal, S.~Roy, G.~Srinivasan, and K.~Roy, ``Proposal for a
  leaky-integrate-fire spiking neuron based on magnetoelectric switching of
  ferromagnets,'' \emph{IEEE Transactions on Electron Devices}, vol.~64, no.~4,
  pp. 1818--1824, 2017.

\bibitem{sengupta2016magnetic}
A.~Sengupta, P.~Panda, P.~Wijesinghe, Y.~Kim, and K.~Roy, ``Magnetic tunnel
  junction mimics stochastic cortical spiking neurons,'' \emph{Scientific
  reports}, vol.~6, no.~1, pp. 1--8, 2016.

\bibitem{bichler2012visual}
O.~Bichler, M.~Suri, D.~Querlioz, D.~Vuillaume, B.~DeSalvo, and C.~Gamrat,
  ``Visual pattern extraction using energy-efficient “2-pcm synapse”
  neuromorphic architecture,'' \emph{IEEE Transactions on Electron Devices},
  vol.~59, no.~8, pp. 2206--2214, 2012.

\bibitem{jo2010nanoscale}
S.~H. Jo, T.~Chang, I.~Ebong, B.~B. Bhadviya, P.~Mazumder, and W.~Lu,
  ``Nanoscale memristor device as synapse in neuromorphic systems,'' \emph{Nano
  letters}, vol.~10, no.~4, pp. 1297--1301, 2010.

\bibitem{nvsim}
X.~Dong, C.~Xu, Y.~Xie, and N.~P. Jouppi, ``Nvsim: A circuit-level performance,
  energy, and area model for emerging nonvolatile memory,'' \emph{IEEE
  Transactions on Computer-Aided Design of Integrated Circuits and Systems},
  vol.~31, no.~7, pp. 994--1007, July 2012.

\bibitem{cite-key}
\BIBentryALTinterwordspacing
H.~Bazzi, A.~Harb, H.~Aziza, M.~Moreau, and A.~Kassem, ``Rram-based
  non-volatile sram cell architectures for ultra-low-power applications,''
  \emph{Analog Integrated Circuits and Signal Processing}, vol. 106, no.~2, pp.
  351--361, 2021. [Online]. Available:
  \url{https://doi.org/10.1007/s10470-020-01587-z}
\BIBentrySTDinterwordspacing

\bibitem{9474064}
C.~Münch and M.~B. Tahoori, ``Testing resistive memory based neuromorphic
  architectures using reference trimming,'' in \emph{2021 Design, Automation
  Test in Europe Conference Exhibition (DATE)}, 2021, pp. 1592--1595.

\bibitem{9365820}
E.~Karl, S.~Shiratake, and J.~Chang, ``Session 24 overview: Advanced embedded
  memories,'' in \emph{2021 IEEE International Solid- State Circuits Conference
  (ISSCC)}, vol.~64, 2021, pp. 332--333.

\bibitem{huai2008spin}
Y.~Huai \emph{et~al.}, ``Spin-transfer torque mram (stt-mram): Challenges and
  prospects,'' \emph{AAPPS bulletin}, vol.~18, no.~6, pp. 33--40, 2008.

\bibitem{khvalkovskiy2013basic}
A.~Khvalkovskiy, D.~Apalkov, S.~Watts, R.~Chepulskii, R.~Beach, A.~Ong,
  X.~Tang, A.~Driskill-Smith, W.~Butler, P.~Visscher \emph{et~al.}, ``Basic
  principles of stt-mram cell operation in memory arrays,'' \emph{Journal of
  Physics D: Applied Physics}, vol.~46, no.~7, p. 074001, 2013.

\bibitem{stt_endur}
J.~J. Kan, C.~Park, C.~Ching, J.~Ahn, Y.~Xie, M.~Pakala, and S.~H. Kang, ``A
  study on practically unlimited endurance of stt-mram,'' \emph{IEEE
  Transactions on Electron Devices}, vol.~64, no.~9, pp. 3639--3646, Sept 2017.

\bibitem{srinivasan2016magnetic}
G.~Srinivasan, A.~Sengupta, and K.~Roy, ``Magnetic tunnel junction based
  long-term short-term stochastic synapse for a spiking neural network with
  on-chip {STDP} learning,'' \emph{Scientific Reports}, vol.~6, no.~1, jul
  2016.

\bibitem{doi:10.1098/rsta.2019.0157}
\BIBentryALTinterwordspacing
I.~Chakraborty, A.~Agrawal, A.~Jaiswal, G.~Srinivasan, and K.~Roy, ``<i>in
  situ</i> unsupervised learning using stochastic switching in magneto-electric
  magnetic tunnel junctions,'' \emph{Philosophical Transactions of the Royal
  Society A: Mathematical, Physical and Engineering Sciences}, vol. 378, no.
  2164, p. 20190157, 2020. [Online]. Available:
  \url{https://royalsocietypublishing.org/doi/abs/10.1098/rsta.2019.0157}
\BIBentrySTDinterwordspacing

\bibitem{doi:10.1146/annurev.neuro.31.060407.125639}
\BIBentryALTinterwordspacing
N.~Caporale and Y.~Dan, ``Spike timing–dependent plasticity: A hebbian
  learning rule,'' \emph{Annual Review of Neuroscience}, vol.~31, no.~1, pp.
  25--46, 2008, pMID: 18275283. [Online]. Available:
  \url{https://doi.org/10.1146/annurev.neuro.31.060407.125639}
\BIBentrySTDinterwordspacing

\bibitem{DAMOUR2015514}
\BIBentryALTinterwordspacing
J.~D’amour and R.~Froemke, ``Inhibitory and excitatory spike-timing-dependent
  plasticity in the auditory cortex,'' \emph{Neuron}, vol.~86, no.~2, pp.
  514--528, 2015. [Online]. Available:
  \url{https://www.sciencedirect.com/science/article/pii/S089662731500210X}
\BIBentrySTDinterwordspacing

\bibitem{devolder2008single}
T.~Devolder, J.~Hayakawa, K.~Ito, H.~Takahashi, S.~Ikeda, P.~Crozat,
  N.~Zerounian, J.-V. Kim, C.~Chappert, and H.~Ohno, ``Single-shot
  time-resolved measurements of nanosecond-scale spin-transfer induced
  switching: Stochastic versus deterministic aspects,'' \emph{Physical review
  letters}, vol. 100, no.~5, p. 057206, 2008.

\bibitem{jaiswal2016comprehensive}
A.~Jaiswal, X.~Fong, and K.~Roy, ``Comprehensive scaling analysis of current
  induced switching in magnetic memories based on in-plane and perpendicular
  anisotropies,'' \emph{IEEE Journal on Emerging and Selected Topics in
  Circuits and Systems}, vol.~6, no.~2, pp. 120--133, 2016.

\bibitem{Mart_nez_Huerta_2013}
\BIBentryALTinterwordspacing
J.~M. Mart{\'{\i}}nez-Huerta, J.~D. L.~T. Medina, L.~Piraux, and A.~Encinas,
  ``Configuration dependent demagnetizing field in assemblies of interacting
  magnetic particles,'' \emph{Journal of Physics: Condensed Matter}, vol.~25,
  no.~22, p. 226003, may 2013. [Online]. Available:
  \url{https://doi.org/10.1088/0953-8984/25/22/226003}
\BIBentrySTDinterwordspacing

\bibitem{doi:10.1063/1.3670002}
\BIBentryALTinterwordspacing
L.~Xu and S.~Zhang, ``Electric field control of interface magnetic
  anisotropy,'' \emph{Journal of Applied Physics}, vol. 111, no.~7, p. 07C501,
  2012. [Online]. Available: \url{https://doi.org/10.1063/1.3670002}
\BIBentrySTDinterwordspacing

\bibitem{papusoi2008probing}
C.~Papusoi, R.~Sousa, J.~Herault, I.~Prejbeanu, and B.~Dieny, ``Probing fast
  heating in magnetic tunnel junction structures with exchange bias,''
  \emph{New Journal of Physics}, vol.~10, no.~10, p. 103006, 2008.

\bibitem{dayan2001theoretical}
P.~Dayan, L.~F. Abbott \emph{et~al.}, ``Theoretical neuroscience (vol. 806),''
  2001.

\end{thebibliography}

\end{document}